\documentclass{article}
\usepackage{spconf}
\usepackage{epsfig,amsmath,amssymb}
\newtheorem{prop}{Proposition}

\title{NEW CRITERIA FOR ITERATIVE DECODING}
\name{F. Alberge, Z. Naja, P. Duhamel}
\address{Laboratoire des Signaux et Syst{\` e}mes,
Univ Paris-Sud, CNRS,\\
Sup{\' e}lec, Plateau de Moulon, 91190 Gif-sur-Yvette, FRANCE\\
email :alberge,naja,pierre.duhamel@lss.supelec.fr }
\begin{document}
\ninept
\maketitle
\normalfont
\begin{abstract}
Iterative decoding was not originally introduced as the solution to an optimization problem rendering the analysis of its convergence very difficult. In this paper, we investigate the link between iterative decoding and classical optimization techniques. We first show that iterative decoding can be rephrased as two embedded minimization processes involving the Fermi-Dirac distance. Based on this new formulation, an hybrid proximal point algorithm is first derived with the additional advantage of decreasing a desired criterion. In a second part, an hybrid minimum entropy algorithm is proposed with improved performance compared to the classical iterative decoding. Even if this paper focus on iterative decoding for BICM, the results can be applied to the large class of turbo-like decoders.
\end{abstract}
\begin{keywords}
Optimization methods, Iterative methods, Decoding.
\end{keywords}

\section{Introduction}
\label{sec:intro}
Bit-Interleaved Coded Modulation (BICM) was first suggested by Zehavi in \cite{Zehavi} to improve the Trellis Coded Modulation performance over Rayleigh-fading channels. In BICM, the diversity order is increased by using bit-interleavers instead of symbol interleavers. This improvement is achieved at the expense of a reduced minimum Euclidean distance leading to a degradation over non-fading Gaussian channels \cite{Zehavi}. This drawback can be overcome by using iterative decoding (BICM-ID) at the receiver \cite{Li2}. BICM-ID is known to provide excellent performance for both Gaussian and fading channels.\\
The iterative decoding scheme used in BICM-ID is very similar to serially concatenated turbo-decoders. Indeed, the serial turbo-decoder makes use of an exchange of information between computationally efficient decoders for each of the component codes. In BICM-ID, the inner decoder is replaced by demapping which is less computationally demanding than a decoding step. Even if this paper focuses on iterative decoding for BICM, the results can be applied to the large class of iterative decoders including serial or parallel concatenated turbo decoders as long as low-density parity-check (LDPC) decoders. Among the different attempts to provide an analysis of iterative decoding, the EXIT chart analysis and density evolution have permitted to make significant progress \cite{ten2,Gamal} but the results developed within this setting apply only in the case of large block length. Another tool of analysis is the connection of iterative decoding to factor graphs \cite{Kschischang} and belief propagation \cite{Pearl}. Convergence results for belief propagation exist but are limited to the case where the corresponding graph is a tree which does not include turbo code or LDPC. A link between iterative decoding and classical optimization algorithms has been made recently in \cite{Walsh} where the turbo decoding is interpreted as a nonlinear block Gauss Seidel iteration. In parallel, a geometrical approach has been considered and provides an interesting interpretation in terms of projections. The particular case of BICM-decoding has been studied in \cite{Muquet,Alberge}. In \cite{Richardson}, the turbo-decoding is interpreted in a geometric setting as a dynamical system leading to new but incomplete results.\\
In this paper we reformulate the iterative decoding as two embedded proximal point algorithms involving the Bregman divergence built on the Fermi-Dirac energy. We prove that each iteration of the decoding decreases a certain criterion.We also propose an hybrid minimum entropy algorithm with improved performance compared to the classical BICM.

\section{BICM-ID with soft decision feedback}
\begin{figure}[!h]
\centerline{\epsfxsize=8cm\epsfysize=1cm\epsfbox{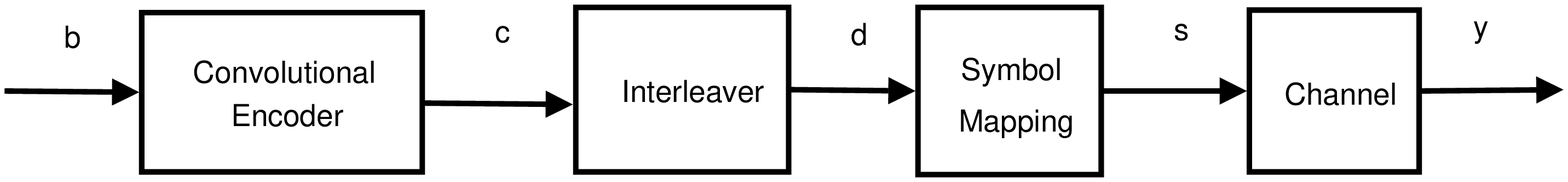}}\vspace*{-0.4cm}\caption{\footnotesize \label{fig_transmitter}
Transmission model}
\end{figure}

\begin{figure}[!h]
\centerline{\epsfxsize=7cm\epsfysize=1.5cm\epsfbox{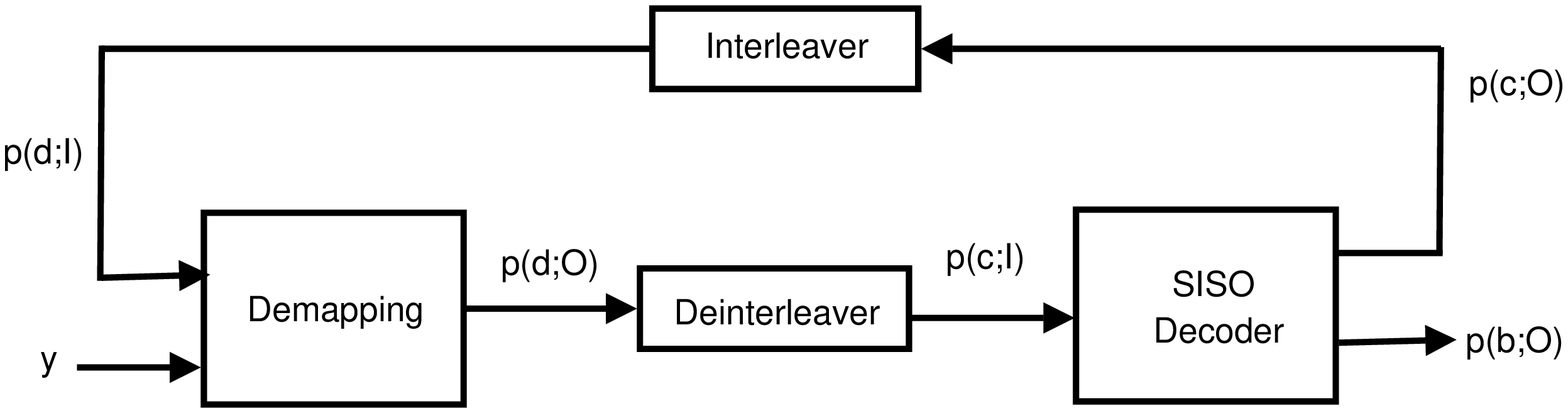}}\vspace*{-0.4cm}\caption{\footnotesize \label{fig_receiver}
Receiver for a BICM-ID with soft-decision feedback}
\end{figure}

A conventional BICM system \cite{Caire} is built from a serial concatenation of a convolutional encoder, a bit interleaver and an M-ary bits-to-symbol mapping (where $M=2^m$) as shown in fig. \ref{fig_transmitter}. The sequence of information bits \underline{\bf b} is first encoded by a convolutional encoder to produce the output encoded bit sequence \underline{\bf c} of length $L_c$ which is then scrambled by a bit interleaver (as opposed to the channel symbols in the symbol-interleaved coded sequence) operating on bit indexes. Let \underline{\bf d} denote the interleaved sequence. Then, $m$ consecutive bits of \underline{\bf d} are grouped as a channel symbol ${\bf d}_k=(d_{km+1},...d_{(k+1)m})$. The complex transmitted signal ${\bf s}_k=\epsilon({\bf d}_k)$ is then chosen from an M-ary constellation $\Psi$ where $\epsilon$ denotes the mapping scheme. For simplicity, we consider transmission over the AWGN channel. The received signal reads:
\begin{equation}
\label{model}
{\bf y}_k={\bf s}_k+{\bf n}_k \;\;\; 1\leq k\leq L_c/m
\end{equation}
where ${\bf n}_k$ is a complex white Gaussian noise with independent in-phase and quadrature components having two-sided power spectral density $\sigma_c^2$.\\
Due to the presence of the random bit interleaver, the true maximum likelihood decoding of BICM is too complicated to implement in practice. Figure \ref{fig_receiver} shows the block diagram of the receiver for a BICM-ID system with soft-decision feedback.

In the first iteration, the encoded bits are assumed equally likely. The demapping consists in evaluating {\em a posteriori} probabilities (APP) for the encoded bits without accounting for the code structure, namely:
\begin{eqnarray}
p_{APP}(d_{km+i}=b)&\sim\sum_{{\bf s}:{\bf s}_k\in\Psi^{i}_{b}}p({\bf y}\vert {\bf s})p({\bf s})\label{APP_map1}\\
&\sim\sum_{{\bf s}_k\in\Psi^{i}_{b}}p({\bf y}_k\vert {\bf s}_k)p({\bf s}_k)\label{APP_map11}
\end{eqnarray}
where ${\bf s}=\{{\bf s}_1,...,{\bf s}_{L_c/m}\}$, ${\bf y}=\{{\bf y}_1,...,{\bf y}_{L_c/m}\}$ and $\Psi^{i}_{b}$, $b\in\{0,1\}$, denotes the subset of $\Psi$ that contains all symbols whose labels have the value $b$ in the $i^{th}$ position. In the turbo decoding process, the quantities exchanged through the blocks are not {\em a posteriori} probabilities (APP) but extrinsic information \cite{Glavieux}. The extrinsic information at the output of the demapping $p(d_{km+i};O)$ is computed as $p_{APP}(d_{km+i})/p(d_{km+i};I)$ where $p(d_{km+i};I)$ is the {\em a priori} information for the demapping sub-block. Since the bit interleaver makes the bits independent, the extrinsic information $p(d_{km+i};O)$ reads:
\begin{equation}
\label{ext_map}
p(d_{km+i}=b;O) = K_m \sum_{{\bf s}_k\in\Psi^{i}_{b}}p({\bf y}_k\vert{\bf s}_k) \prod_{j\neq i}p(d_{km+j};I)
\end{equation}
and the corresponding APP reads:
\begin{equation}
\label{app_map}
p_{APP}(d_{km+i}=b)= K'_m \sum_{{\bf s}_k\in\Psi^{i}_{b}}p({\bf y}_k\vert{\bf s}_k) \prod_{j}p(d_{km+j};I)
\end{equation}
where $K_m$ and $K'_m$ are normalization factors. The extrinsic information $p(d_{km+i};O)$ is de-interleaved and delivered to the SISO decoder \cite{Benedetto} as an {\em a priori} information on the encoded bits. Let $c_l=d_{\sigma^{-1}(km+j)}$ where $\sigma^{-1}$ is for the permutation on the indexes due to the deinterleaver; $p(c_l;I)$ is the updated input of the Single Input Single Output (SISO) decoder. The extrinsic information at the output of the SISO decoder is obtained through \cite{Muquet,Moher}:
\begin{equation}
\label{ext_siso}
p(c_l=b;O) = K_c \sum_{{\bf c}\in{\cal R}^{l}_{b}}{\bf I}_{\cal C}({\bf c})\prod_{j\neq l}p(c_j;I)
\end{equation}
and the corresponding APP is:
\begin{equation}
\label{APP_siso}
p_{APP}(c_l=b)= K_c' \sum_{{\bf c}\in{\cal R}^{l}_{b}}{\bf I}_{\cal C}({\bf c})\prod_{j}p(c_j;I)
\end{equation}
where ${\bf I}_{\cal C}({\bf c})$ stands for the indicator function of the code, i.e. ${\bf I}_{\cal C}({\bf c})=1$ if ${\bf c}$ is a codeword and $0$ otherwise and ${\cal R}^{l}_{b}$ denotes the set of binary words of length $L_c$ with value $b$ in the $l^{th}$ position. $K_c$ and $K_c'$ are normalization factors. The extrinsic information $p(c_l;O)$ is interleaved and delivered to the demapping sub-block as a regenerated {\em a priori} information. If the process converges the APP of the two sub-blocks are the same. The criteria proposed in the following are based on this property and encourage a faster convergence towards this objective.\\

\section{Notations from Information Geometry}
\subsection{Basic tools}
 We first introduce some notations that will be useful in the sequel. Let ${\bf B}_i \in \{0,1\}^N$ denote the binary representation of the integer $i, 0\leq i\leq 2^{N-1}$. The binary representation of all the words of length N is gathered into matrix ${\bf B} = ({\bf B}_0,{\bf B}_1,...,{\bf B}_{2^N-1})^T$ with dimension $2^N\times N$. Let $\bf{\eta}$ be a probability mass function on the outcomes $\chi={\bf B}_i$ then 
$$\bf{\eta}=(Pr[\chi={\bf B_0}],Pr[\chi=\bf{ B_1}],...,Pr[\chi=\bf{ B_{2^N-1}}])^T$$
Given a PMF $\eta$, its log-coordinates are the vector ${\bf \theta}$ whose $i^{th}$ element is given by $\theta_i=\ln (Pr[\chi={\bf B_{i}}])-\ln (Pr[\chi={\bf  B_{0}}])$. We can observe that there is a one-to-one mapping between $\bf{\eta}$ and $\bf{\theta}$ since the vector  $\bf{\eta}$ can be written $\bf{\eta} = exp({\bf \theta}-\psi({\bf \theta}))$ where $\psi({\bf \theta})=log(\sum_i exp((\theta)_i))$.We also introduce the bitwise log-probability ratios with elements of the form $\lambda_j=log(\frac{Pr[\chi_j=1]}{Pr[\chi_j=0]})$ where $\chi_j$ is the $j^{th}$ bit of the binary word $\chi$ and ${\bf \lambda}\in{\mathbb R}^N$. For factorisable probability measures ({\em ie} PMF that factors into the product of their bitwise marginals so that $Pr(\chi)=\Pi_jPr(\chi_j)$), the log-coordinates take the form ${\bf \theta}={\bf B}{\bf \lambda}$. 

\subsection{Link with iterative decoding}
Let ${\bf \theta}_m$ denote the log-coordinates vector of the PMF $p({\bf y}\vert{\bf s})$. Let ${\bf \lambda_1}$ denote the log-probability ratio 
corresponding to the prior $p(d_{km+i};I)$ such that:
$$({\lambda_1})_{km+i}=\ln\left(\frac{p(d_{km+i}=1;I)}{p(d_{km+i}=0;I)}\right)$$
Thus, the log-coordinates of $p({\bf y}\vert{\bf s})\Pi_{j,k}p(d_{km+i};I)$ reads ${\bf B\lambda_1+\theta_m}$. \\
Let $p_{\bf B\lambda_1+\theta_m}$ represent the vector whose $i^{th}$ element is the probability that the $i^{th}$ bit is $1$ according 
to the measure with log-coordinate ${\bf B\lambda_1+\theta_m}$. The APP at the output of the demapper merge with $p_{\bf B\lambda_1+\theta_m}$. From eq. (\ref{ext_map})-(\ref{app_map}), $p_{APP}(d_{km+i}=b)=
p(d_{km+i}=b;I)p(d_{km+i}=b;0)$, the log-coordinates of the APP at the output of the demapper also merge with ${\bf B(\lambda_1+\lambda_2)}$ where 
$({\lambda_2})_{km+i}=\ln\left(\frac{p(d_{km+i}=1;O)}{p(d_{km+i}=0;O)}\right)$. Then the demapper sub-block solves, with respect to ${\bf \lambda_2}$, the equation:
\begin{equation}\label{eq_mapp}
p_{\bf B(\lambda_1+\lambda_2)} = p_{\bf B\lambda_1+\theta_m}
\end{equation}
Let ${\bf \theta}_c$ denote the log-coordinates of the PMF associated with the indicator function. Then the decoder sub-block solves, with respect to ${\bf \lambda_1}$, the equation:
\vspace*{-0.3cm}
\begin{equation}\label{eq_cod}
p_{\bf B(\lambda_1+\lambda_2)} = p_{\bf B\lambda_2+\theta_c}
\end{equation}
Iterative decoding is thus equivalent to:
$$\begin{array}{l}
find \; {\bf \lambda_2}^{(k+1)} \; such \; that \;\; p_{\bf B(\lambda_1^{(k)}+\lambda_2^{(k+1)})} = p_{\bf B\lambda_1^{(k)}+\theta_m}\\
find \; {\bf \lambda_1}^{(k+1)} \; such \; that \;\; p_{\bf B(\lambda_1^{(k+1)}+\lambda_2^{(k+1)})} = p_{\bf B\lambda_2^{(k+1)}+\theta_c}\\
\end{array}
$$
At the convergence, the APP from the two sub-blocks should be in accordance {\em ie} $ p_{\bf B(\lambda_1^{(\infty)}+\lambda_2^{(\infty)})} = p_{\bf B\lambda_1^{(\infty)}+\theta_m}=p_{\bf B\lambda_2^{(\infty)}+\theta_c}$.

\section{An optimization problem}
The Fermi-Dirac divergence is the Bregman divergence built on the Fermi-Dirac entropy $f({\bf p}) = \sum_j p_j\ln (p_j)+(1-p_j)\ln (1-p_j)$ with $dom(f)=[0;1]$. The Fermi-Dirac divergence reads 
$$D_{FD}({\bf p},{\bf q})= \sum_j p_j \ln\left(\frac{p_j}{q_j}\right) + \sum_j (1-p_j)\ln\left(\frac{1-p_j}{1-q_j}\right)$$
and is exactly the Kullback-Leibler distance for bit probabilities. The Fermi-Dirac divergence is a non-symmetric distance. As we can notice, this distance is very convenient for computing distances between bit probabilities.
\begin{prop}
The demapping sub-block solves the minimization problem
\vspace*{-0.3cm}
$$\min_{\bf \lambda_2}D_{FD}(p_{\bf B\lambda_1+\theta_m},p_{\bf B(\lambda_1+\lambda_2)})$$
The decoding sub-block solves the minimization problem
$$\min_{\bf \lambda_1}D_{FD}(p_{\bf B\lambda_2+\theta_c},p_{\bf B(\lambda_1+\lambda_2)})$$
\end{prop} 
{\em Proof: 
The proof is obvious by noting that $(\lambda_1+\lambda_2)_{km+i} = \ln\left(\frac{p_{B(\lambda_1+\lambda_2)}(d_{km+i}=1)}{p_{B(\lambda_1+\lambda_2)}(d_{km+i}=0)}\right)$ thus $p_{\bf B(\lambda_1+\lambda_2)}=\frac{exp({\bf \lambda_1+\lambda_2})}{1+exp({\bf \lambda_1+\lambda_2})}$}.\\
This proposition illustrates that iterative decoding can be formulated as two embedded minimization steps based on the Fermi-Dirac distance. In the next section, we investigate some modifications of this original criterion.

\subsection{An hybrid proximal point algorithm}
In the classical iterative decoding, the two minimization steps seem independent meaning that the minimization of one of the criterion does 
not imply necessarily a decrease of the other criterion at the next iteration. Proximal point methods \cite{Luque} permit to make the link between the two criteria.
These methods are generally used to guarantee the monotonicity of the convergence process often at the cost of a slow convergence speed. 
Following the proximal point technique we obtain the minimization process:
\begin{eqnarray}
{\bf \lambda_2^{(k+1)}} =\min_{\bf \lambda_2} J_{\theta_m}({\bf \lambda_1},{\bf \lambda_2} )= \min_{\bf \lambda_2} D_{FD}({\bf p_{B\lambda_1+\theta_m}},{\bf p_{B(\lambda_1+\lambda_2)}})\nonumber\\
+\mu_m D_{FD}({\bf p_{B(\lambda_1^{(k)}+\lambda_2^{(k)})}},{\bf p_{B(\lambda_1+\lambda_2)}})\nonumber\\
{\bf \lambda_1^{(k+1)}} =\min_{\bf \lambda_1} J_{\theta_c}({\bf \lambda_1},{\bf \lambda_2} )=\min_{\bf \lambda_1} D_{FD}({\bf p_{B\lambda_2+\theta_c}},{\bf p_{B(\lambda_1+\lambda_2)}})\nonumber\\
+\mu_c D_{FD}({\bf p_{B(\lambda_1^{(k)}+\lambda_2^{(k+1)})}},{\bf p_{B(\lambda_1+\lambda_2)}})\nonumber
\end{eqnarray}
As can be seen, the original criterion is modified through the addition of a penalization term in order to encourage smooth variations of the successive estimates.
This minimization process is equivalent to finding ${\bf \lambda_2^{(k+1)}}$ such that 
\begin{equation}
{\bf p_{B(\lambda_1^{(k)}+\lambda_2^{(k+1)})}} = \frac{{\bf p_{B\lambda_1^{(k)}+\theta_m}}+\mu_m{\bf p_{B(\lambda_1^{(k)}+\lambda_2^{(k)})}}}{1+\mu_m}
\label{l2_pp}
\end{equation}
and  ${\bf \lambda_1^{(k+1)}} $ such that 
\begin{equation}{\bf p_{B(\lambda_1^{(k+1)}+\lambda_2^{(k+1)})}} = \frac{\bf p_{B\lambda_2^{(k+1)}+\theta_c}+\mu_c{\bf p_{B(\lambda_1^{(k)}+\lambda_2^{(k+1)})}}}{1+\mu_c}
\label{l1_pp}
\end{equation}
Note that this new procedure also converges towards solutions satisfying (\ref{eq_mapp}) and (\ref{eq_cod}). A good choice for the parameters $\mu_m$ and $\mu_c$ permits to ensure that each criterion decreases with the iterations. Actually, we want to enforce $J_{\theta_m}({\bf \lambda_1^{(k)}},{\bf \lambda_2^{(k+1)}} )\leq J_{\theta_c}({\bf \lambda_1^{(k)}},{\bf \lambda_2^{(k)}} )$. Since the Fermi-Dirac distance is convex with respect to its second parameter, we have $J_{\theta_m}({\bf \lambda_1^{(k)}},{\bf \lambda_2^{(k+1)}} )\leq \frac{\mu_m}{1+\mu_m}(D_{FD}({\bf p_{B\lambda_1^{(k)}+\theta_m}},{\bf p_{B(\lambda_1^{(k)}+\lambda_2^{(k)})}})+D_{FD}({\bf p_{B(\lambda_1^{(k)}+\lambda_2^{(k)})}},{\bf p_{B\lambda_1^{(k)}+\theta_m}}))$. Moreover, we also have $ D_{FD}({\bf p_{B\lambda_2^{(k)}+\theta_c}},{\bf p_{B(\lambda_1^{(k)}+\lambda_2^{(k)})}})\leq J_{\theta_c}({\bf \lambda_1^{(k)}},{\bf \lambda_2^{(k)}})$. Connecting the two relations, we obtain an upper bound for $\mu_m$:
$$\mu_m\leq\frac{D_{FD}({\bf p_{B\lambda_2^{(k)}+\theta_c}},{\bf p_{B(\lambda_1^{(k)}+\lambda_2^{(k)})}})}{{\cal D}_{FD}-D_{FD}({\bf p_{B\lambda_2^{(k)}+\theta_c}},{\bf p_{B(\lambda_1^{(k)}+\lambda_2^{(k)})}})}$$
where ${\cal D}_{FD}$ is a symmetric distance, namely ${\cal D}_{FD}= D_{FD}({\bf p_{B\lambda_1^{(k)}+\theta_m}},{\bf p_{B(\lambda_1^{(k)}+\lambda_2^{(k)})}})+D_{FD}({\bf p_{B(\lambda_1^{(k)}+\lambda_2^{(k)})}},{\bf p_{B\lambda_1^{(k)}+\theta_m}})$. The upper bound for $\mu_c$ can be obtained in the same way. Iterating (\ref{l2_pp}) and (\ref{l1_pp}) with $\mu_c$ and $\mu_m$ correctly chosen yields an algorithm that converges towards the same points than the classical iterative decoding with the additional advantage of decreasing at each iteration a desired criterion. In the next section, we propose a new criterion in order to improve the performance of the iterative decoding.

\subsection{An hybrid minimum entropy algorithm}
The entropy of the vectors of marginals ${\bf p_{B(\lambda_1+\lambda_2)}}$ is defined as
\begin{eqnarray}
E_{B(\lambda_1+\lambda_2)}=-\sum_n p_{B(\lambda_1+\lambda_2)}(n)log2(p_{B(\lambda_1+\lambda_2)}(n))\nonumber\\
-\sum_n(1-p_{B(\lambda_1+\lambda_2)}(n))log2(1-p_{B(\lambda_1+\lambda_2)}(n))\nonumber
\end{eqnarray}
The quantity $E_{B(\lambda_1+\lambda_2)}$ gives a measure of the reliability of the decisions. Indeed, $E_{B(\lambda_1+\lambda_2)}\rightarrow 0$ does not always mean that the decisions are correct, but rather that the iterative decoding algorithm is confident about its decisions. Nevertheless, in the iterative decoding, the decisions are in most cases correct when $E_{B(\lambda_1+\lambda_2)}\rightarrow 0$ \cite{Kocarev}. In this section, we propose a new criterion that minimizes $E_{B(\lambda_1+\lambda_2)}$ under the constraint $D_{FD}({\bf p_{B\lambda_1+\theta_m}},{\bf p_{B(\lambda_1+\lambda_2)}})\leq \epsilon$ for the demapping and $D_{FD}({\bf p_{B\lambda_2+\theta_c}},{\bf p_{B(\lambda_1+\lambda_2)}})$ for the decoding. This is equivalent to:
\begin{eqnarray}
{\bf \lambda_2^{(k+1)}}=\min_{\bf \lambda_2} D_{FD}({\bf p_{B\lambda_1+\theta_m}},{\bf p_{B(\lambda_1+\lambda_2)}})+\eta_m E_{B(\lambda_1+\lambda_2)}\label{E1}\\
{\bf \lambda_1^{(k+1)}}=\min_{\bf \lambda_1} D_{FD}({\bf p_{B\lambda_2+\theta_c}},{\bf p_{B(\lambda_1+\lambda_2)}})+\eta_c E_{B(\lambda_1+\lambda_2)}\label{E2}
\end{eqnarray}
By zeroing the gradient of the two criteria in (\ref{E1}) and (\ref{E2}), we obtain the new update equations:
\begin{eqnarray}
{\bf \lambda_2}^{(k+1)}: & f_{\eta_m}(p_{B(\lambda_1^{(k)}+\lambda_2^{(k+1)})}(n))=p_{B\lambda_1^{(k)}+\theta_m}(n)\;\; 1\leq n\leq L_c\nonumber\\
{\bf \lambda_1}^{(k+1)}: & f_{\eta_c}(p_{B(\lambda_1^{(k+1)}+\lambda_2^{(k+1)})}(n))=p_{B\lambda_2^{(k+1)}+\theta_c}(n)\;\; 1\leq n\leq L_c\nonumber
\end{eqnarray}
where $f_{\eta}(p_{B(\lambda_1+\lambda_2)}(n))=p_{B(\lambda_1+\lambda_2)}(n)-\eta p_{B(\lambda_1+\lambda_2)}(n)(1-p_{B(\lambda_1+\lambda_2)}(n))\log\left(\frac{p_{B(\lambda_1+\lambda_2)}(n)}{1-p_{B(\lambda_1+\lambda_2)}(n)}\right)$. 
\begin{figure}[!h]
\centerline{\epsfig{figure=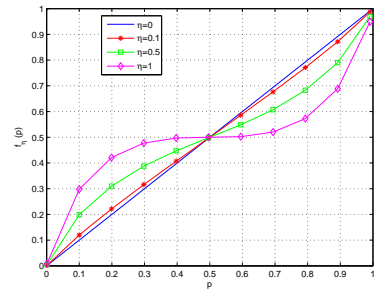,width=2.2in,angle=0}}
\caption{\label{distorsion}
$f_\eta(p)$ for various values of $\eta$}
\end{figure}
The function $f_\eta$ is plotted on fig (\ref{distorsion}). We can notice that: {\em (i)} the distortion increases with $\eta$ {\em (ii)} $f_{\eta}(p)$ belongs to $[0;1]$ {\em (iii)} $f_{\eta}(p)$ is a strictly increasing function. As a consequence each step of the minimization process has a unique solution that can be found using classical techniques.

\section{Simulation}
\begin{figure}[!h]
\centerline{\epsfig{figure=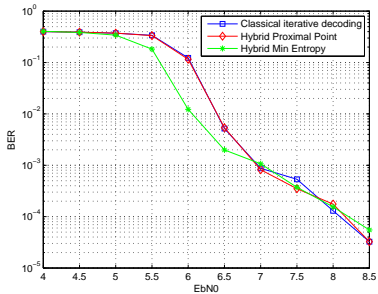,width=2.2in,angle=0}}
\vspace*{-0.4cm}
\caption{\label{result_BER}
BER versus EbN0}
\end{figure}

\begin{figure}[!h]
\centerline{\epsfig{figure=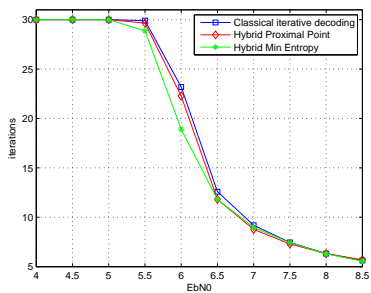,width=2.2in,angle=0}}
\vspace*{-0.4cm}
\caption{\label{result_iter}
Iteration number versus EbN0}
\end{figure}

We compare the performance in terms of bit error rate and iteration number of the classical iterative decoding with the hybrid proximal point algorithm (HPP) and also with the hybrid minimum entropy algorithm (HMEA). Each algorithm stops when the Fermi-Dirac distance between the APP of the two sub-blocks is less than  $10^{-3}$ or when $30$ iterations are reached. The generator polynomial of the encoder is $g=[1 1 1 ; 0 0 1 ;1 0 0 ]$. The bits are mapped using subset partitioning to a 8-PSK modulation. The length of the coded bit sequence is $L_c=6000$. The step-sizes $\eta_m$ and $\eta_c$ in the HMEA are both chosen equal to $0.05$. The results are plotted in fig. (\ref{result_BER}) and (\ref{result_iter}). We can see that the classical iterative decoding and the HPP exhibits exactly the same performance. This is not surprising concerning the BER since both methods converge towards the same points. We can also notice that these results are obtained with the same number of iterations in both cases meaning that the proximal point technique does not reduce, in this case, the convergence speed. Both methods have almost the same computational complexity with the additional advantage for the HPP to minimize a desired criterion with the iterations. As expected, the HMEA outperforms the others methods in terms of BER in the middle area with a number of iterations at most equal to the number of iterations needed in the classical BICM. However, this last method has a higher computational complexity due to the distortion function $f_\eta$.
\vspace*{-0.7cm}
\section{Conclusion}
In this paper, iterative decoding is rephrased as two embedded minimization processes. From this formulation, we have derived an hybrid proximal point algorithm that exhibits the same performance than the classical iterative decoding. This proximal point algorithm decreases at each step a well identified criterion. We have also built an hybrid minimum entropy algorithm. The minimization of the entropy leads to an improvement of the performance. 

\bibliographystyle{IEEEbib}
\bibliography{bib_Icassp08}

\begin{thebibliography}{10}

\bibitem{Zehavi}
E.~Zehavi,
\newblock ``{8-PSK trellis codes for a Rayleigh fading channel},''
\newblock {\em IEEE Trans. Commun.}, vol. 40, pp. 873--883, May 1992.

\bibitem{Li2}
X.~Li, A.~Chindapol, and J.A. Ritcey,
\newblock ``Bit interleaved coded modulation with iterative decoding and
  8-{PSK} signaling,''
\newblock {\em IEEE trans Commun.}, vol. 50, pp. 1250--1257, Aug 2002.

\bibitem{ten2}
S.~ten Brink,
\newblock ``Convergence behavior of iteratively decoded parallel concatenated
  codes,''
\newblock {\em IEEE trans Commun.}, vol. 49, pp. 1727--1737, Oct 2001.

\bibitem{Gamal}
H.~El Gamal and A.R. Hammons,
\newblock ``Analysing the turbo decoder using the {G}aussian approximation,''
\newblock {\em IEEE Trans. on Inform. Theory}, vol. 47, pp. 671--686, Feb.
  2001.

\bibitem{Kschischang}
F.R. Kschischang, B.J. Frey, and H.A. Loeliger,
\newblock ``Factor graphs and the sum-product algorithm,''
\newblock {\em IEEE Trans. on Inform. Theory}, vol. 47, pp. 498--519, Feb.
  2001.

\bibitem{Pearl}
J.~Pearl,
\newblock {\em Probabilistic {Reasoning in Intelligent Systems: Network of
  Plausible Inference}},
\newblock San Francisco, CA: Morgan Kaufmann, 1988.

\bibitem{Walsh}
J.~M. Walsh, P.A. Regalia, and C.~R. Johnson,
\newblock ``{Turbo decoding as Iterative Constrained Maximum-Likelihood
  Sequence Detection},''
\newblock {\em IEEE Trans. on Inform. Theory}, vol. 52, pp. 5426--5437, Dec.
  2006.

\bibitem{Muquet}
B.~Muquet, P.~Duhamel, and M.~de~Courville,
\newblock ``A geometrical interpretation of iterative turbo decoding,''
\newblock in {\em Proc. Int. Symposium on Inform. Theory}, Lausanne,
  Switzerland, May 2002.

\bibitem{Alberge}
F.~Alberge,
\newblock ``{Iterative decoding as Dykstra's algorithm with alternate
  I-projection and reverse I-projection},''
\newblock in {\em EUSIPCO Proc.}, Lausanne, Switzerland, August 2008.

\bibitem{Richardson}
T.~Richardson,
\newblock ``The geometry of turbo-decoding dynamics,''
\newblock {\em IEEE Trans. on Inform. Theory}, vol. 46, no. 1, pp. 9--23, 2000.

\bibitem{Caire}
G.~Caire, G.Taricco, and E.~Biglieri,
\newblock ``Bit-interleaved coded modulation,''
\newblock {\em IEEE Trans. on Inform. Theory}, vol. 4, pp. 927--946, May 1998.

\bibitem{Glavieux}
C.~Berrou, A.~Glavieux, and P.~Thitimajshima,
\newblock ``Near {Shannon} limit error-correcting coding and decoding: Turbo
  codes,''
\newblock in {\em Proc. IEEE Int. Conf. Commun}, 1993, pp. 1064--1070.

\bibitem{Benedetto}
S.~Benedetto, D.~Divsalar, G.~Montorsi, and F.~Pollara,
\newblock ``A soft-input soft-output {APP} module for iterative decoding of
  concatenated codes,''
\newblock {\em IEEE Commun. Letters}, vol. 1, pp. 22--24, Jan 1997.

\bibitem{Moher}
M.~Moher and T.A. Gulliver,
\newblock ``Cross-entropy and iterative decoding,''
\newblock {\em IEEE Trans. on Inform. Theory}, vol. 44, no. 7, pp. 3097--3104,
  Nov. 1998.

\bibitem{Luque}
F.J. Luque,
\newblock ``Asymptotic convergence analysis of the proximal point algorithm,''
\newblock {\em SIAM Journal on Control and Optimization}, vol. 22, no. 2, pp.
  277--293, 1984.

\bibitem{Kocarev}
L.~Kocarev, F.~Lehmann, G.M. Maggio, B.~Scanavino, Z.~Tasev, and A.~Vardy,
\newblock ``Nonlinear dynamics of iterative decoding systems: analysis and
  applications,''
\newblock {\em IEEE Trans. on Infor. Theory}, vol. 52, no. 4, pp. 1366--1384,
  2006.

\end{thebibliography}
\end{document}